\documentclass[journal=jacsat,manuscript=article]{achemso}

\usepackage[version=3]{mhchem} % Formula subscripts using \ce{}

\author{Akshay A. Murthy}
\affiliation{Superconducting Quantum Materials and Systems Division, Fermi National Accelerator Laboratory (FNAL), Batavia, IL 60510, USA}
\email{amurthy@fnal.gov}
\author{Paul Masih Das}
\affiliation{Department of Materials Science and Engineering, Northwestern University, Evanston, IL, 60208 USA}
\author{Stephanie M. Ribet}
\affiliation{Department of Materials Science and Engineering, Northwestern University, Evanston, IL, 60208 USA}
\alsoaffiliation{International Institute of Nanotechnology, Northwestern University, Evanston, IL, 60208 USA}
\author{Cameron Kopas}
\affiliation{Rigetti Computing, Berkeley, CA 94710, USA}
\author{Jaeyel Lee}
\affiliation{Superconducting Quantum Materials and Systems Division, Fermi National Accelerator Laboratory (FNAL), Batavia, IL 60510, USA}
\author{Matthew J. Reagor}
\affiliation{Rigetti Computing, Berkeley, CA 94710, USA}
\author{Lin Zhou}
\affiliation{Ames Laboratory, U.S. Department of Energy, Ames, IA 50011, United States}
\author{Matthew J. Kramer}
\affiliation{Ames Laboratory, U.S. Department of Energy, Ames, IA 50011, United States}
\author{Mark C. Hersam}
\affiliation{Department of Materials Science and Engineering, Northwestern University, Evanston, IL, 60208 USA}
\alsoaffiliation{Department of Chemistry, Northwestern University, Evanston, IL 60208}
\alsoaffiliation{Department of Electrical and Computer Engineering, Northwestern University, Evanston, IL 60208}
\author{Mattia Checchin}
\affiliation{Superconducting Quantum Materials and Systems Division, Fermi National Accelerator Laboratory (FNAL), Batavia, IL 60510, USA}
\author{Anna Grasselino}
\affiliation{Superconducting Quantum Materials and Systems Division, Fermi National Accelerator Laboratory (FNAL), Batavia, IL 60510, USA}
\author{Roberto dos Reis}
\affiliation{Department of Materials Science and Engineering, Northwestern University, Evanston, IL, 60208 USA}
\alsoaffiliation{The NU\textit{ANCE} Center, Northwestern University, Evanston, IL, 60208 USA}
\author{Vinayak P. Dravid}
\affiliation{Department of Materials Science and Engineering, Northwestern University, Evanston, IL, 60208 USA}
\alsoaffiliation{The NU\textit{ANCE} Center, Northwestern University, Evanston, IL, 60208 USA}
\alsoaffiliation{International Institute of Nanotechnology, Northwestern University, Evanston, IL, 60208 USA}
\email{v-dravid@northwestern.edu}
\author{Alexander Romanenko}
\affiliation{Superconducting Quantum Materials and Systems Division, Fermi National Accelerator Laboratory (FNAL), Batavia, IL 60510, USA}
\email{aroman@fnal.gov}

\title[An \textsf{achemso} demo]
 {Developing a Chemical and Structural Understanding of the Surface Oxide in a Niobium Superconducting Qubit}

\abbreviations{}

\begin{document}

\begin{abstract}
Superconducting thin films of niobium have been extensively employed in transmon qubit architectures. Although these architectures have demonstrated remarkable improvements in recent years, further improvements in performance through materials engineering will aid in large-scale deployment. Here, we use information retrieved from secondary ion mass spectrometry and electron microscopy to conduct a detailed assessment of the surface oxide that forms in ambient conditions for transmon test qubit devices patterned from a niobium film. We observe that this oxide exhibits a varying stoichiometry with NbO and NbO$_2$ found closer to the niobium film and Nb$_2$O$_5$ found closer to the surface. In terms of structural analysis, we find that the Nb$_2$O$_5$ region is semicrystalline in nature and exhibits randomly oriented grains on the order of 1-2 nm corresponding to monoclinic N-Nb$_2$O$_5$ that are dispersed throughout an amorphous matrix. Using fluctuation electron microscopy, we are able to map the relative crystallinity in the Nb$_2$O$_5$ region with nanometer spatial resolution. Through this correlative method, we observe that amorphous regions are more likely to contain oxygen vacancies and exhibit weaker bonds between the niobium and oxygen atoms. Based on these findings, we expect that oxygen vacancies likely serve as a decoherence mechanism in quantum systems.
\end{abstract}

\noindent KEYWORDS: superconducting qubits, 4D-STEM, electron diffraction,  Nb thin films, interfaces, hydrides, decoherence mechanisms
\\~\\

% \keywords{\textit{In situ} electron microscopy, heterostructure, transition metal dichalcogenides, MoS$_{2}$, differential phase contrast}

\section{Introduction}
Over the last two decades, many significant advances have been made towards constructing large-scale quantum computers. In particular, superconducting quantum information technology has emerged as a leading architecture to interrogate complex problems commonly deemed intractable with the most efficient classical computing platforms \cite{doi:10.1146/annurev-conmatphys-031119-050605, Wendin_2017, deLeoneabb2823, murray2021material}. Nonetheless, extending this technology to large-scale devices requires continued progress to improve reliability and performance. Such improvements require higher quality materials and specifically, an increased understanding and control over imperfections including interfaces and surfaces \cite{doi:10.1063/5.0079321}.

As an example, in the case of niobium (Nb)-superconducting transmon qubits, the amorphous surface oxides that form upon ambient exposure serve as major sources of microwave dissipation. At milliKelvin (mK) operating temperatures, they also display loss tangent values that are three orders of magnitude larger compared to the Nb thin films and Si substrates \cite{PhysRevApplied.13.034032, doi:10.1063/5.0017378, PhysRevApplied.16.014018}. This loss can be largely attributed to two-state defects in the amorphous surface oxide, i.e. two-level system (TLS) defects \cite{PhysRevApplied.13.034032}. These states emerge as a result of deviations from long-range order. The system is able to transition between the two states through low energy excitations at the operating temperatures for superconducting transmon qubit devices \cite{premkumar2020microscopic, towards2019}. While the density of these TLS can be calculated through spectroscopic methods \cite{RN21} and several TLS noise mechanisms have been proposed (tunneling of atoms,\cite{PhysRevLett.95.210503} tunneling of electrons,\cite{PhysRevB.87.144201} paramagnetic spins\cite{RevModPhys.86.361}, quasiparticle traps \cite{doi:10.1126/sciadv.abc5055}), an understanding of the atomic defects or nanoscale deviations from which this noise fundamentally arises remains unclear. 

A combination of various spectroscopy and microscopy techniques have led researchers to understand that deviations from crystalline order on the nanoscale introduce quantum decoherence \cite{towards2019}. As such, scanning transmission electron microscopy (STEM) is an indispensable tool for identification of materials that host TLS defects in a broader effort to eliminate these sources. Namely, the ability to procure a variety of analytical and spectroscopic signals from nanometric volumes allows for detection of chemical, structural, and electromagnetic fluctuations on these relevant length scales \cite{ophus_2019,  RIBET2021}.  We employ a combination of electron microscopy and time-of-flight secondary ion mass spectrometry (TOF-SIMS) to identify nanoscale defects and chemical inhomogeneities within the surface oxide associated with a transmon test qubit device. The use of cutting edge techniques to map the relative crystallinity allows us to link the presence of sub-stoichiometric oxygen concentrations to local disorder, which suggests that oxygen vacancies may serve as a decoherence mechanism in niobium-based quantum systems. 

\section{Results and Discussion}
Nb transmon test qubits were fabricated following the procedures detailed by Nersisyan et al. \cite{8993458} This involved preparation of a Si (001) wafer (float-zone $>$10,000 Ohm-cm) with an RCA surface treatment \cite{doi:10.1063/1.3517252, 8993458}, followed by deposition of Nb films via high-power impulse magnetron sputtering (HiPIMS) with a base pressure less than 1E-8 Torr at room temperature. TOF-SIMS measurements were performed using a dual beam system (IONTOF 5) to analyze the concentration and depth distribution of oxygen in the Nb film. Secondary ion measurements were performed using a liquid bismuth ion beam (Bi+). A cesium ion gun with an energy of 500 eV was used for sputtering the surface for depth profile measurements to detect anions. TEM samples were prepared from the Nb contact pad using a 30 kV focused Ga$^+$ ion beam. We first provide an assessment of the chemical composition of the Nb surface oxide before discussing nanoscale inhomogeneities in this region using advanced electron microscopy techniques.

\subsection{Surface Oxide Chemical Composition}

\begin{figure}
\includegraphics[width=7in]{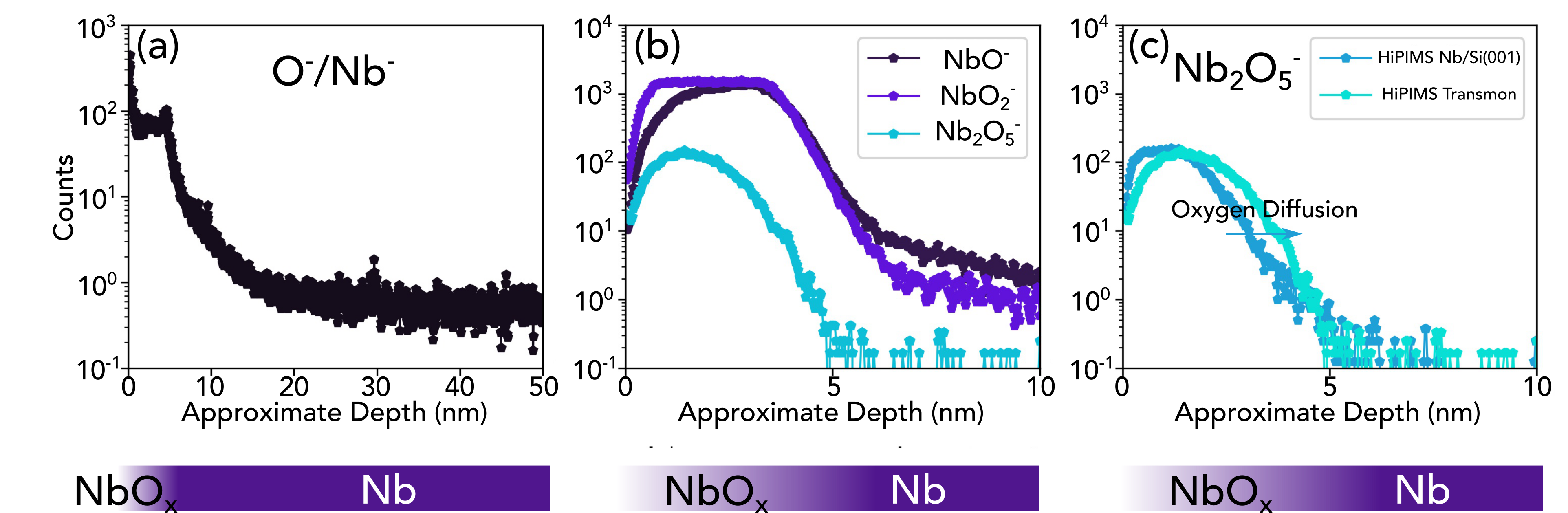}
\caption{TOF-SIMS depth profiles taken from Nb thin films. (a) Normalized oxygen signal is plotted as a function of depth taken from a Nb pad in a transmon test qubit. (b) Different NbO$_x$ signals are plotted as a function of depth for the top 10 nm of a Nb pad in a transmon test qubit. We observe that the Nb$_2$O$_5$ signal peaks closer to the surface whereas the NbO$_2$ and NbO signals persist further into the film. (c) Nb$_2$O$_5$  signal is plotted as a function of depth taken from a Nb pad in a transmon test qubit and an unprocessed Nb thin film deposited with HiPIMS. We find that the lithography and etching processes performed following film deposition lead to inward diffusion of oxygen.}
\label{fgr:TOF_SIMS}
\end{figure}

TOF-SIMS depth profiles were taken from transmon test qubit devices and the averaged profiles are provided in Figure \ref{fgr:TOF_SIMS}. TOF-SIMS is a valuable method for examining impurities such as oxygen in these films as it combines sensitivity to light elements and high mass resolution with $<$100 nm spatial resolution. From these measurements (Figure \ref{fgr:TOF_SIMS}a), we observe an appreciable content of oxygen in the immediate vicinity of the surface, which corresponds to the surface oxide as discussed previously \cite{doi:10.1063/5.0079321}. We also observe that this oxygen signal decays to a baseline level of roughly 0.5 O$^-$/Nb$^-$ counts, which we expect results from the incorporation of oxygen interstitial atoms in the Nb film \cite{PhysRevB.9.888}.

Further, TOF-SIMS enables the identification of different oxide species based on their mass to charge (m/z) ratios. While direct quantification of these oxides can be challenging due to the possibility for molecular fragmentation prior to detection by the analyzer, monitoring these signals provides qualitative information regarding their spatial location within the oxide layer \cite{BOSE2020145464}. In Figure \ref{fgr:TOF_SIMS}b, we observe that the signal associated with Nb$_2$O$_5$ exhibits a peak closer to the surface whereas the NbO$_2$ and NbO signals persist further into the film. This is in agreement with the literature indicating a change in valence state for the Nb throughout the oxide \cite{doi:10.1063/1.1663849, BOSE2020145464}. As the presence of Nb$_2$O$_5$ has been specifically linked to decoherence in Nb superconducting qubit systems \cite{PhysRevApplied.13.034032}, we focus our analysis to this region throughout the text. 

To get a better understanding for how this oxide evolves with qubit device processing, we compare the Nb$_2$O$_5$ signal taken from Nb thin films prior to the lithography and etching processes required to fabricate transmon qubits as well as that taken from similar processed test qubit devices. The averaged signals for each are plotted in Figure \ref{fgr:TOF_SIMS}c. We observe that device processing steps following film deposition lead to oxygen diffusion into the Nb film. Such an effect has been previously observed in Nb superconducting radiofrequency (SRF) cavities \cite{https://doi.org/10.48550/arxiv.2108.13352}. In that situation, the authors observed that this oxygen diffusion led to a significant degradation in performance and was likely caused by an increased oxygen vacancy concentration. As the two research areas are intrinsically linked, observations in the SRF cavity arena can help inform methods to improve superconducting qubit performance metrics as well.

\subsection{Structural and Chemical Analysis}

\begin{figure*}
\includegraphics[width=3.5in]{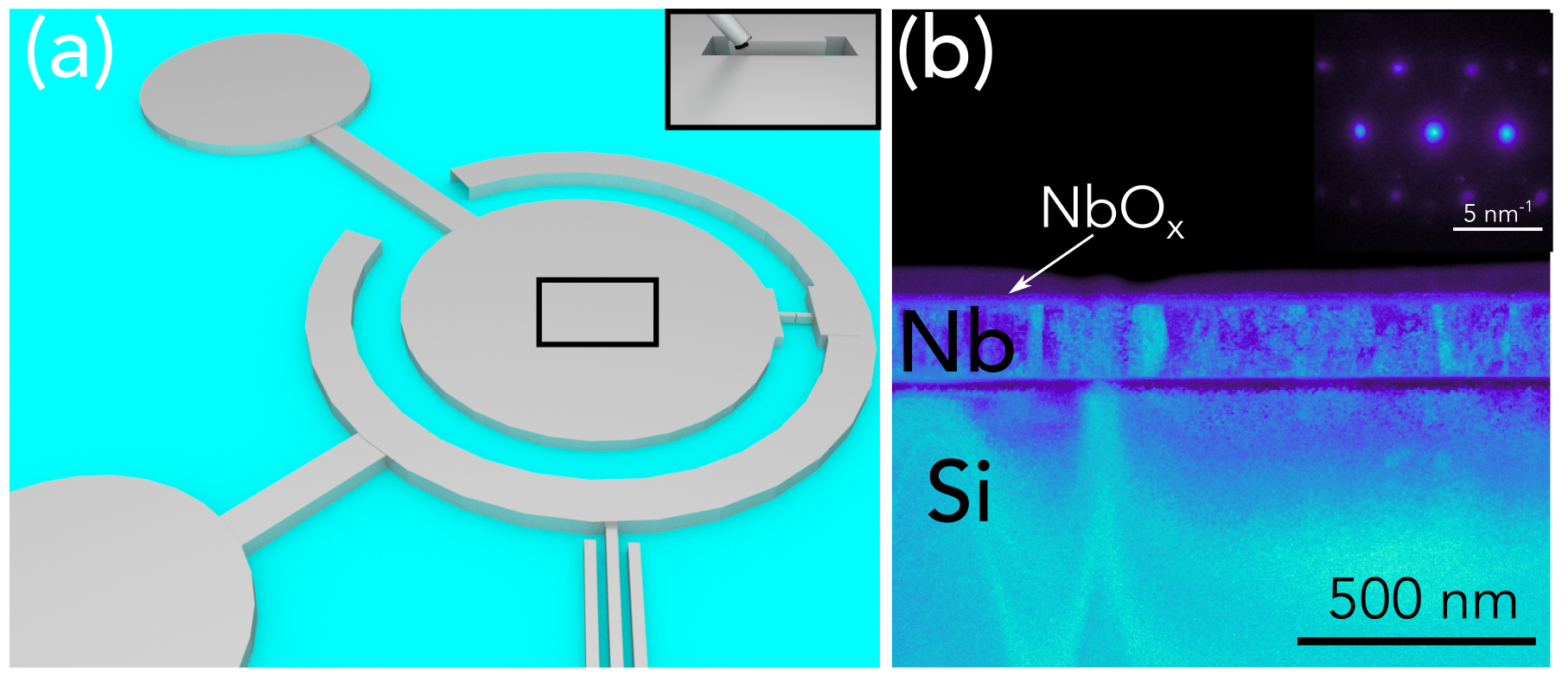}
\caption{Electron microscopy of transmon test qubit (a) Schematic of transmon qubit architecture. Black box represents the Nb contact pad region investigated in this study. (b) Annular dark field image taken from a representative, cross-sectional TEM sample. The surface oxide is indicated. A diffraction pattern taken from the Nb metal film is provided as an inset.}
\label{fgr:schematic_low_mag}
\end{figure*}

To probe the surface oxide further, we performed electron microscopy and diffraction analysis on a cross-section taken from the contact pad of the transmon test qubit (Figure \ref{fgr:schematic_low_mag}a). An annular dark field (ADF) image of this TEM lamella using collection angles varying between 10 mrad to 70 mrad is provided in Figure \ref{fgr:schematic_low_mag}b with the NbO$_x$ region labeled. From this image, we observe that the Nb film exhibits columnar grains that are roughly 50 nm by 170 nm and are characteristic of the HiPIMS process. A diffraction pattern taken when the Nb film was tilted along the [111] zone axis is provided as an inset.

\begin{figure*}
\includegraphics[width=7in]{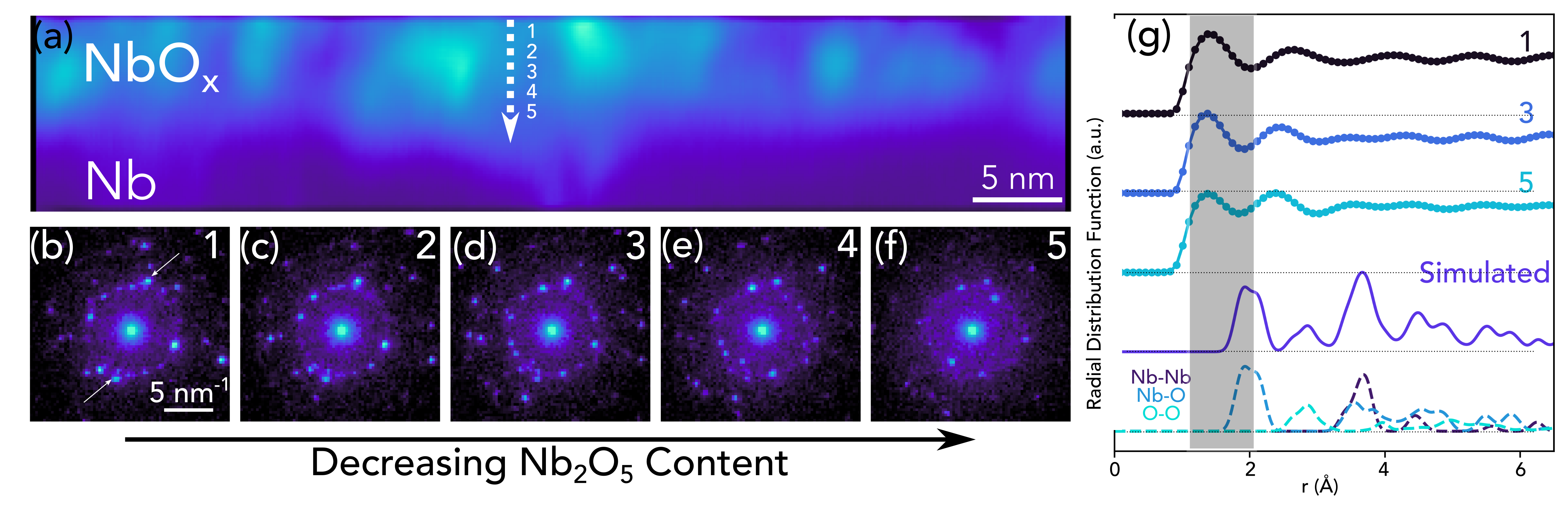}
\caption{Nb/O interface (a) DF image of Nb oxide/Nb constructed using a virtual detector that matched the diffraction ring of Nb oxide seen in Figure S2b. (b-f) Diffraction patterns taken from regions 1, 2, 3, 4, and 5, respectively. The distinct diffraction spots indicated by the arrows are associated with N-Nb$_2$O$_5$ as indicated by the simulation provided in Figure S3. As these spots fade in intensity from region 1 to region 5, it is apparent that the Nb$_2$O$_5$ content decreases in the direction away from the surface. (g) Comparison between experimental RDF as a function of position and simulated RDF for crystalline Nb$_2$O$_5$. This simulated profile is also decomposed into a pair distribution function to understand which pairs of atoms give rise to the features observed.}
\label{fgr:Nb_oxide}
\end{figure*}

To clearly identify the structural and chemical features associated with the surface oxide, a dark field STEM image of this area and the underlying Nb thin film is provided in Figure \textbf{\ref{fgr:Nb_oxide}}a. This image was constructed by applying a virtual detector with an inner collection angle of 10 mrad and an outer collection angle of 15 mrad, which was designed specifically to match the initial diffraction ring observed in the NbO$_x$ diffraction pattern (Figure S2). Representative diffraction patterns captured from the indicated regions in the dark field image in Figure \ref{fgr:Nb_oxide}a are provided in Figure \ref{fgr:Nb_oxide}d-h. We observe a combination of intense diffraction spots as well as broad diffuse rings. Based on comparison with a simulated diffraction pattern, we conclude that these spots likely arise from N-Nb$_2$O$_5$, which exhibits a monoclinic structure (Figure S3) \cite{NICO20161}. These spots are present for spots 1-4 in Figure \textbf{\ref{fgr:Nb_oxide}}a, which suggests that this area corresponds to Nb$_2$O$_5$. Conversely, the broad diffuse rings are indicative of a lack of long-range order. The appearance of both of these features suggests that Nb$_2$O$_5$ is semi-crystalline in nature and once again we focus our analysis to this region.

To gain a qualitative understanding of the size of the crystallites in the Nb$_2$O$_5$, we implement virtual apertures, or binary masks, to each captured diffraction pattern in order to selectively filter signal associated with a few of the most prominent diffraction spots. This use of virtual apertures produces the dark-field image in Figure S3. Based on this image, it appears that the N-Nb$_2$O$_5$ crystallites are roughly around 1-3 nm in dimension and distributed throughout the entire oxide layer.

To better understand the nature of the amorphous regions within the Nb$_2$O$_5$, we need to employ specialized methods as conventional diffraction and analysis techniques are insufficient in providing nanoscale structural information from these regions. Instead fluctuation electron microscopy (FEM) allows us to better understand the medium range ordering present in the sample. This analysis was conducted following the approaches detailed previously. \cite{doi:10.1063/5.0015532, doi:10.1063/5.0069549, MU20161, savitzky_2021}. First, this process involves using a 1 nm probe and capturing diffraction patterns as a function of position. Following elliptical correction, the radial intensity was then calculated as a function of scattering angle. By subtracting the contribution associated with the thermal motion of atoms as well as the scattering resulting from individual atoms, we were able to calculate the scattering factor as a function of position. Finally upon performing a Fourier transform, the radial distribution function (RDF) as a function of real-space distance is accessible. An RDF profile represents the probability of finding an atom $r$ distance away from a reference atom. As such, atomic bond distances will produce peaks in the profile.

\begin{figure*}
\includegraphics[width=7in]{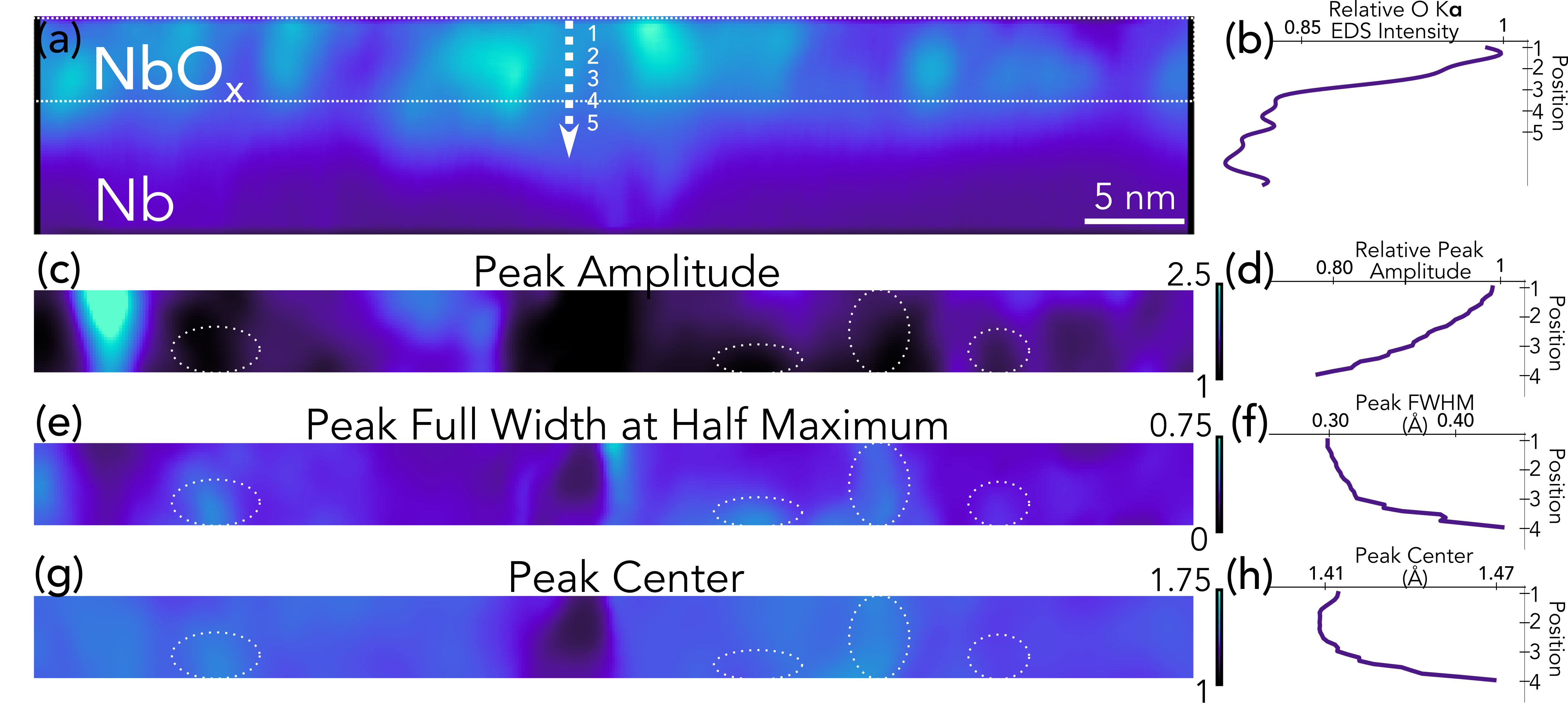}
\caption{Fluctuation Electron Microscopy of Surface Oxide (a) DF image of Nb oxide/Nb constructed using a virtual detector. (b) Line profile of EDS intensity as a function of position. (c-h) 2D maps and averaged 1D profiles of various peak parameters for Nb-O bond through the Nb$_2$O$_5$ region. Peak amplitude values are provided in c-d. Peak FWHM values are provided in e-f. Peak center values are provided in g-h. Regions with a relatively large peak FWHM values are indicative of relatively disordered regions. A few of these regions are circled in (e). Based on the associated peak center and peak amplitude maps, it appears that disordered regions in the Nb$_2$O$_5$ are more likely to contain reduced oxygen concentrations and larger bond distances between Nb and O.}
\label{fgr:Nb_RDF}
\end{figure*}

The RDF calculated for positions 1, 3, and 5 in the dark field image are seen in Figure \ref{fgr:Nb_oxide}c. From these profiles, we observe that the most prominent feature is centered about 1.8 Å. Based on the calculated radial distribution function profile for crystalline Nb$_2$O$_5$ (Figure S1) \cite{Farrow_2007}, this primary feature is expected to arise from Nb-O bonding. As the experimental RDF is obtained from a semicrystalline region, disorder arising from local deviations in the bond length and bond angles introduce the experimental peak broadening observed compared to the simulated profile. Nonetheless, we can monitor how this feature evolves throughout the oxide to obtain information associated with the nature of the Nb-O bonds in this region.

To verify the assignment of this feature in the RDF profile to the Nb-O bond, we map and provide the averaged line profiles of various parameters associated with this peak in Figure \ref{fgr:Nb_RDF} for the Nb$_2$O$_5$ region. Specifically, the parameters of the Nb-O peak provide information about the distribution of Nb-O bonds as a function of position. For instance, in addition to the peak amplitude representing the local density of Nb-O bonds, the peak full width at half maximum (FWHM) provides insight into the degree of local order or disorder, and the peak center provides information regarding the local bond distances. The Nb-O feature was fitted using a Gaussian distribution (details are provided in SI). From Figure \ref{fgr:Nb_RDF}d, we observe that the amplitude of this feature steadily decreases beginning at the surface in the direction towards the Nb film. This is in agreement with the EDS line profile provided in Figure \ref{fgr:Nb_RDF}b as well as the TOF-SIMS data discussed previously. In conjunction with the simulated RDF profiles, this further corroborates that this primary feature arises from Nb-O bonding.

Having established that this initial peak represents the Nb-O bond, we calculate and plot additional peak parameters which provide information on the relative crystallinity and bond distances. With respect to the peak FWHM, we observe that this parameter varies throughout the Nb$_2$O$_5$ region. In Figure \ref{fgr:Nb_oxide}e, regions exhibiting large FWHM values are circled. These areas correspond to regions that are highly disordered and as discussed previously, more prone to hosting TLS. Upon evaluation of the peak amplitudes present in these disordered regions (circled in Figure \ref{fgr:Nb_oxide}c), we find that local disorder tends to be linked to Nb-O peak amplitude suppression, which potentially arises through an increase in oxygen vacancy concentration as suggested by the TOF-SIMS measurements. We find that the cross-correlation R$^2$ value between these two parameters in the Nb$_2$O$_5$ region is 0.29. Similarly, we observe that these disordered regions tend to exhibit an increased peak center as well (Figure \ref{fgr:Nb_oxide}g). This is indicative of weaker bonds between Nb and O and in this case, we find that the R$^2$ value between these two parameters within the Nb$_2$O$_5$ region is roughly 0.22. 

Since TLS likely arise from disordered regions in Nb$_2$O$_5$, the correlation in the peak parameters suggests that in the case of this system, TLS may arise due to the presence of oxygen vacancies. Further, the weak Nb-O bonds in this region may assist in permitting atomic tunneling at low temperatures \cite{PhysRevLett.95.210503}. As theoretical work \cite{sheridan2021microscopic} and experimental findings \cite{5719535} have suggested that oxygen vacancies introduce local paramagnetic moments, such a tunneling process at the qubit operating temperatures is expected to introduce electromagnetic noise and potentially lead to quantum decoherence.

\section{Concluding Remarks}
In this study, we applied TOF-SIMS and STEM imaging and diffraction methods to identify structural inhomogeneities and defects in the surface oxide associated with a transmon test qubit. Using these methods, we observe that this oxide exhibits a varying stoichiometry with NbO and NbO$_2$ found closer to the niobium film and Nb$_2$O$_5$ found closer to the surface. Further, we find that the Nb$_2$O$_5$ region is semicrystalline in nature and exhibits randomly oriented grains on the order of 1-2 nm corresponding to N-Nb$_2$O$_5$ that are dispersed through an amorphous matrix. By subsequently mapping the relative crystallinity in the Nb$_2$O$_5$ region, we observe that amorphous regions are more likely to contain oxygen vacancies and exhibit weaker bonds between the niobium and oxygen atoms. Together, this work suggests that oxygen vacancies in the surface oxide likely serve as a decoherence mechanism in quantum systems. As a result, identifying methodologies for passivation of oxygen vacancies or preventing the formation of this lossy surface oxide are critical to improving coherence times in superconducting qubits.

\section{Methods}

\subsection*{TEM Sample Preparation}
TEM samples were prepared from the Nb electrode using a 30 kV focused Ga$^+$ ion beam. In order to protect the surface oxide during the ion milling process, the sample was first coated with 50 nm of Ni. The samples were finely polished to a thickness of roughly 50 nm using 5 kV and 2 kV Ga$^+$ ions in an effort to remove surface damage and amorphization in the regions of interest.\\~\\

\subsection*{STEM Data Collection}
S/TEM data was collected on JEOL 300F Grand ARM S/TEM using an accelerating voltage of 300kV for the low magnification analysis of the Nb film. In this case, the camera length is set to 80 cm and the condenser aperture is selected to provide a convergence semi-angle of 10 mrad. Four-dimensional STEM (4DSTEM) data sets~\cite{ophus_2019} were acquired in a 360 x 62 mesh (with pixel size of 3.33 nm) across the Nb films using a Gatan® OneView camera and synchronized using STEMx. In the case of the Nb surface oxide, an accelerating voltage of 80kV was selected to reduce beam damage in the amorphous compounds and data acquired using a Gatan® Stela hybrid-pixelated camera. STEM-EDS measurements were collected at 200kV using a JEOL Silicon Drift Detector (SDD) with a solid angle of 1.7 sr.\\~\\

\subsection*{Fluctuation Electron Microscopy} \label{sec:RDF}
A condenser aperture was selected to provide a convergence semi-angle of 1.5 mrad for fluctuation electron microscopy measurements. This yields a probe size that is roughly 1 nm and allows for characterizing amorphous regions that are on the order of 5 nm in dimension. The resultant image was subsequently passed through a Gaussian filter with a filter size of 2 pixels to eliminate scanning distortions. Analysis of scanning diffraction-based fluctuation electron microscopy measurements was performed using the py4DSTEM package \cite{savitzky_2021} per the methods outlined by \citeauthor{SOUZAJUNIOR2021441} \cite{SOUZAJUNIOR2021441}. Upon obtaining diffraction patterns from the regions of interest, the diffraction patterns were fit radially using least squares fitting of the first diffraction ring. The radial intensity was then calculated as a function of scattering angle. By subtracting the contribution associated with the thermal motion of atoms as well as the scattering resulting from individual atoms, we were able to calculate the scattering factor as a function of position as discussed. This contribution was approximated by fitting these functions to the tail of the radial intensity and this process yields the reduced structure factor. Finally upon performing a Fourier transform, the radial distribution function (RDF) as a function of real-space distance can be obtained. The initial peak in the RDF was fit with a Gaussian lineshape. The parameters of this Gaussian function are reported.\\~\\

\begin{acknowledgement}

This material is based upon work supported by the U.S. Department of Energy, Office of Science, National Quantum Information Science Research Centers, Superconducting Quantum Materials and Systems Center (SQMS) under the contract No. DE-AC02-07CH11359. Fermilab is operated by the Fermi Research Alliance, LLC under contract No. DE-AC02-07CH11359 with the United States Department of Energy. This work made use of the EPIC facility of Northwestern University’s NU\textit{ANCE} Center, which received support from the Soft and Hybrid Nanotechnology Experimental (SHyNE) Resource (NSF ECCS-1542205); the MRSEC program (NSF DMR-1720139) at the Materials Research Center; the International Institute for Nanotechnology (IIN); the Keck Foundation; and the State of Illinois, through the IIN. The authors thank members of the Superconducting Quantum Materials and Systems (SQMS) Center for valuable discussion. S. M. R. gratefully acknowledges support from IIN and 3M. The authors thank Dr. Anahita Pakzad from Ametek/Gatan, Inc, Pleasanton, CA, for the valuable feedback on the usage of the Stela pixelated detector.

\end{acknowledgement}

\section{Author Contributions}
The manuscript was written through contributions of all authors. All authors have given approval to the final version of the manuscript.

\section{Competing Financial Interests:} The authors declare no competing financial interests.\\

\begin{suppinfo}

EDS map, evolution of diffraction patterns in Nb oxide along with simulated diffraction patterns associated with Nb$_2$O$_5$ are provided in supporting information.

\end{suppinfo}

\bibliography{achemso-demo}

\newpage
% \subsection{For Table of Contents Only}
% \begin{figure*}
% \includegraphics[height=4cm]{TOC.png}
% \end{figure*}

\newpage
\setcounter{figure}{0}
\setcounter{page}{1}
\setcounter{section}{0}

\renewcommand{\thepage}{S\arabic{page}}
\renewcommand{\thesection}{S\arabic{section}}
\renewcommand{\thefigure}{S\arabic{figure}}
\subsection{Supplementary Information}
\section{Developing a Chemical and Structural Understanding of the Surface Oxide in a Niobium Superconducting Qubit}

% \author{Akshay A. Murthy$^{1*}$}
% \author{Paul Masih Das$^2$, Stephanie M. Ribet$^{2,3}$, Cameron Kopas$^4$, Jaeyel Lee$^1$, Matthew J. Reagor$^4$, Lin Zhou$^5$, Matthew J. Kramer$^5$, Mark C. Hersam$^{2,6,7}$, Mattia Checchin$^1$, Anna Grassellino$^1$, Roberto dos Reis$^{2,3,8}$, Vinayak P. Dravid$^{2,3,8*}$, Alexander Romanenko$^{1}$}
% \email[Correspondence email address: ]{amurthy@fnal.gov, v-dravid@northwestern.edu, aroman@fnal.gov}

% \affiliation{$^1$Superconducting Quantum Materials and Systems Division, Fermi National Accelerator Laboratory (FNAL), Batavia, IL 60510, USA}
% \affiliation{$^2$Department of Materials Science and Engineering, Northwestern University, Evanston, IL 60208, USA}
% \affiliation{$^3$International Institute of Nanotechnology, Northwestern University, Evanston, IL 60208, USA}
% \affiliation{$^4$Rigetti Computing, Berkeley, CA 94710, USA}
% \affiliation{$^5$Ames Laboratory, U.S. Department of Energy, Ames, IA 50011, United States}
% \affiliation{$^6$Department of Chemistry, Northwestern University, Evanston, IL 60208}
% \affiliation{$^7$Department of Electrical and Computer Engineering, Northwestern University, Evanston, IL 60208}
% \affiliation{$^8$The NU\textit{ANCE} Center, Northwestern University, Evanston, IL 60208, USA }

\author{\fontsize{12}{15} \selectfont \noindent Akshay A. Murthy,$^{*,\dagger}$ Paul Masih Das,$^{\ddagger}$ Stephanie M. Ribet,$^{\ddagger,\P}$ Cameron Kopas,$^\S$  Jaeyel Lee,$^\dagger$\\
\fontsize{12}{15} \selectfont Matthew J. Reagor,${^\S}$ Lin Zhou,${^\lVert}$ Matthew J. Kramer,${^\lVert}$ Mark C. Hersam,$^{\ddagger,\perp,\#}$\\
\fontsize{12}{15} \selectfont Mattia Checchin,${^\dagger}$ Anna Grassellino,${^\dagger}$ Roberto dos Reis,$^{\ddagger,@}$ Vinayak P. Dravid,$^{*,\ddagger,@,\P }$} and Alexander Romanenko$^{*,\dagger}$

\date{%
    \fontsize{10}{15} \selectfont \noindent
    $^{\dagger}$Superconducting Quantum Materials and Systems Division, Fermi National Accelerator Laboratory (FNAL), Batavia, IL 60510, USA\\%
    $^{\ddagger}$epartment of Materials Science and Engineering, Northwestern University, Evanston, IL 60208, USA\\
    ${^\P}$International Institute of Nanotechnology, Northwestern University, Evanston, IL 60208, USA\\
    ${^\S}$Rigetti Computing, Berkeley, CA 94710, USA\\
    ${^\lVert}$Ames Laboratory, U.S. Department of Energy, Ames, IA 50011, United States\\
    ${^\perp}$Department of Chemistry, Northwestern University, Evanston, IL 60208\\
    ${^\#}$Department of Electrical and Computer Engineering, Northwestern University, Evanston, IL 60208\\  
    ${^@}$The NU\textit{ANCE} Center, Northwestern University, Evanston, IL 60208, USA \\~\\ 
    
    $^*$Corresponding Authors:\\
    Akshay Murthy: \underline{amurthy@fnal.gov}\\
    Vinayak P. Dravid: \underline{v-dravid@northwestern.edu}\\
    Alexander Romanenko: \underline{aroman@fnal.gov}

}
\noindent KEYWORDS: superconducting qubits, 4D-STEM, electron diffraction,  Nb thin films, interfaces, hydrides, decoherence mechanisms
\\~\\
\noindent \textbf{Table of contents:\\
S1: Energy dispersive spectroscopy map of the Nb surface oxide\\
S2: Diffraction patterns taken from Nb oxide\\
S3: Diffraction patterns taken from Nb/Si interface\\
}

\newpage
\begin{figure}[ht]
\includegraphics[width=3.5in]{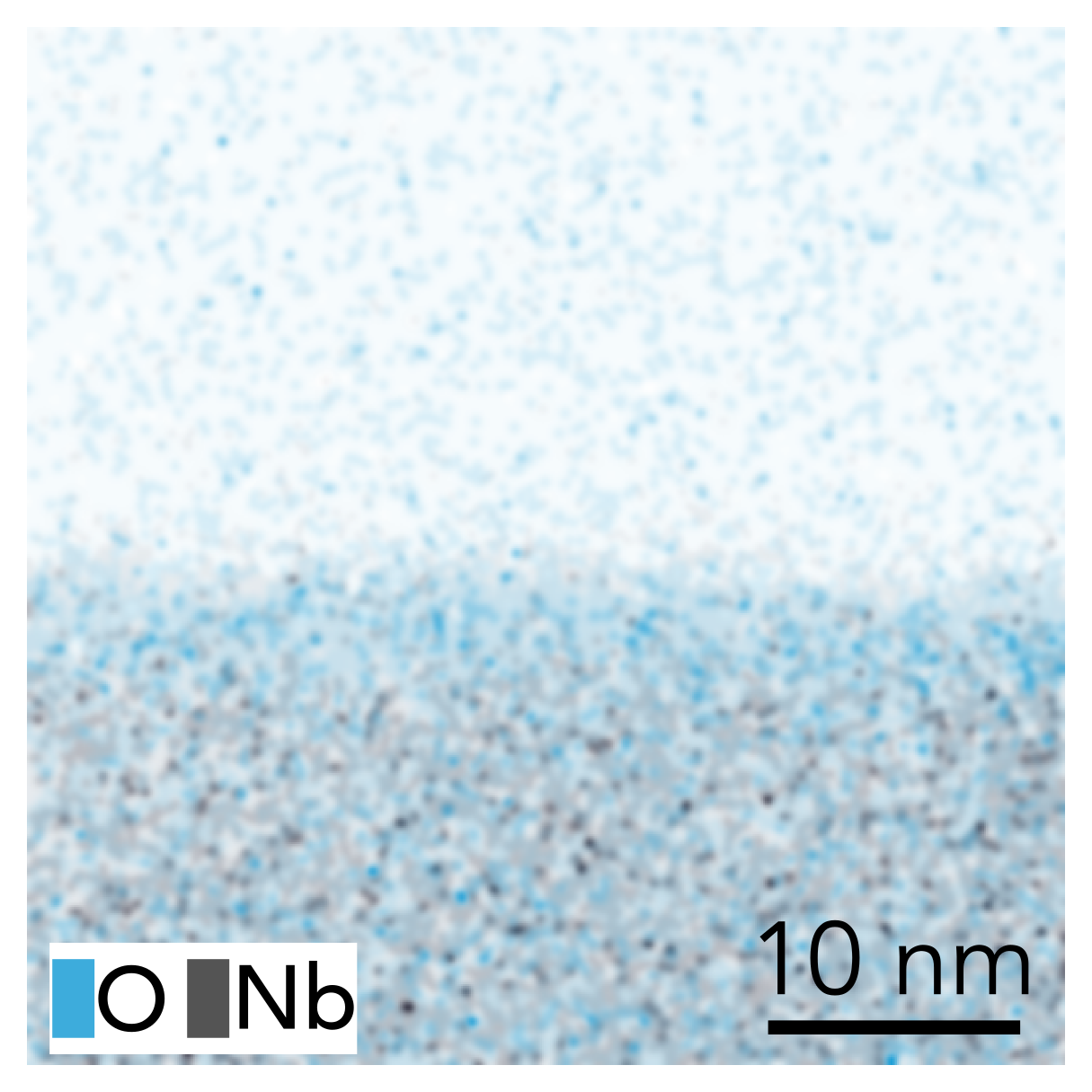}
\caption{Energy dispersive spectroscopy map of the Nb surface oxide.}
\label{fgr:S1}
\end{figure}

\begin{figure*}
\includegraphics[width=7in]{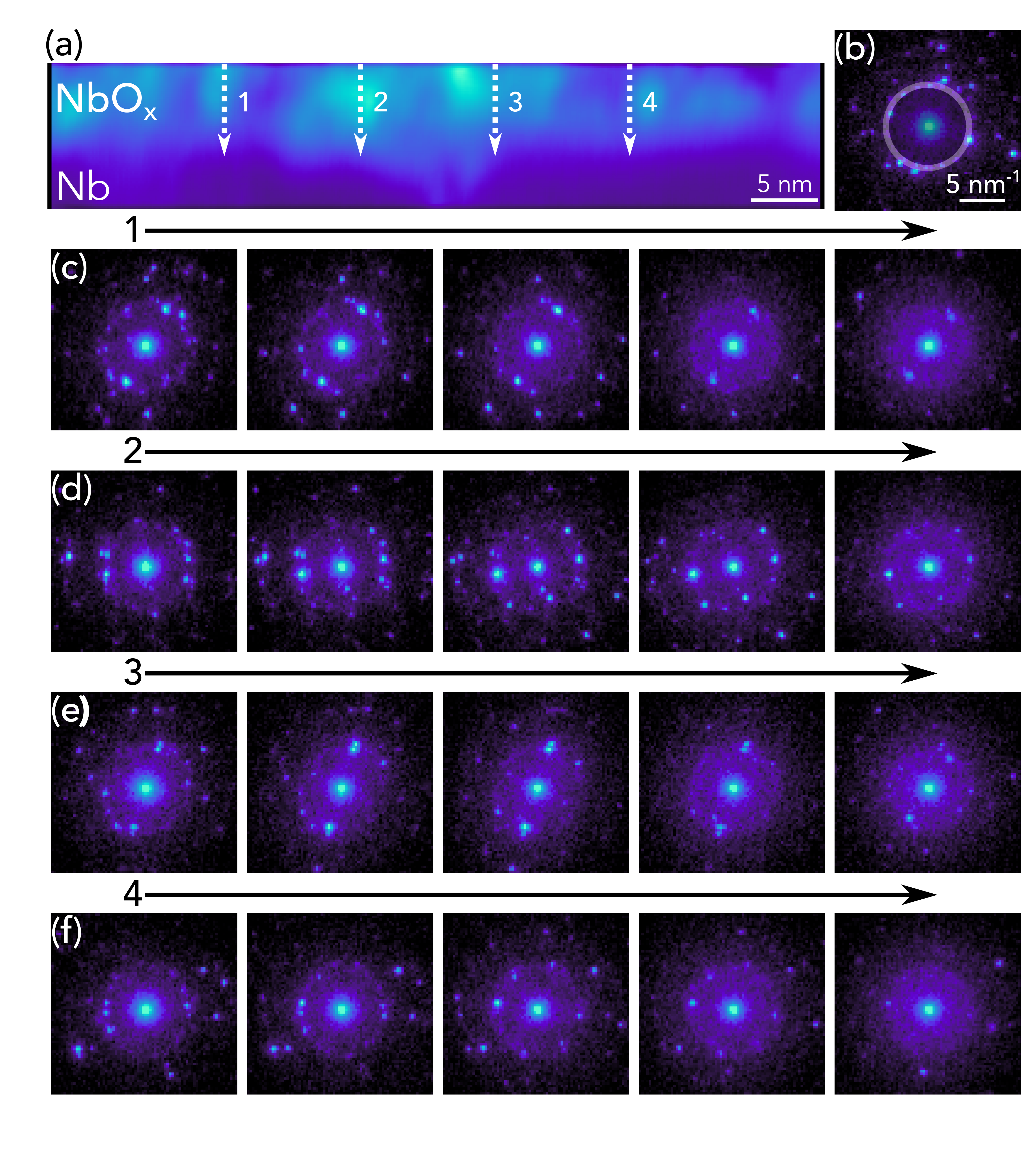}
\caption{(a) DF image of Nb oxide/Nb constructed using a virtual detector that matched the diffraction ring of Nb oxide. The dotted arrows indicate regions over which changes in the diffraction pattern are provided in (c-f). (b) Diffraction pattern taken from NbO$_x$ with a virtual annular detector with an inner collection angle of 10 mrad and an outer collection angle of 15 mrad to preferentially produce a NbO$_x$ dark field image. (c-f) Evolution of the electron diffraction pattern in the direction of the arrows indicated in (a) for the labeled regions 1, 2, 3, 4, and 5, respectively. In all cases, prominent diffraction spots tend to decay in intensity when moving away from the surface.}
\label{fgr:S2}
\end{figure*}

\begin{figure*}
\includegraphics[width=7in]{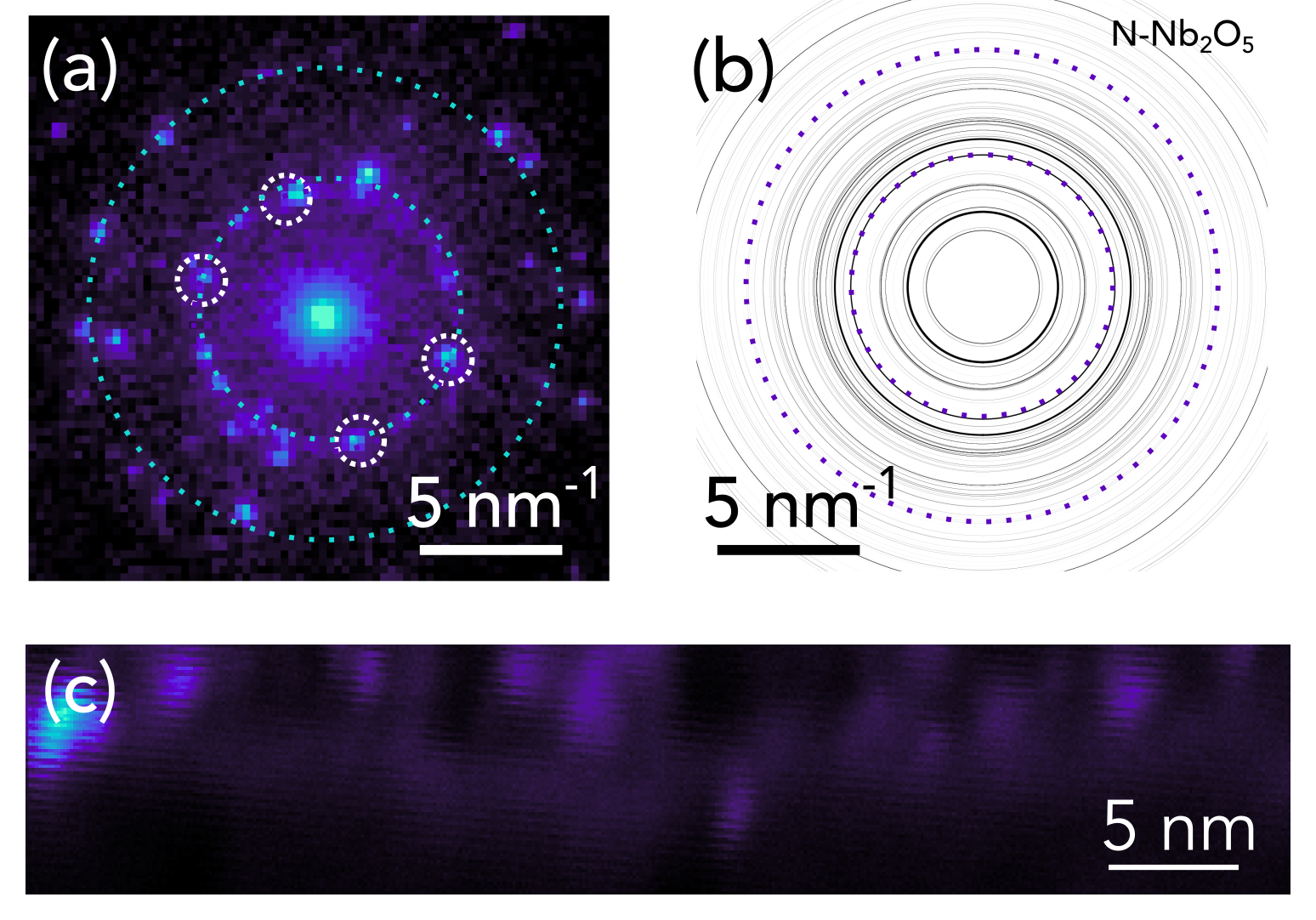}
\caption{(a) Representative diffraction pattern of Nb$_2$O$_5$ taken from region 3 in Figure 3. (b) Simulated diffraction pattern for N-Nb$_2$O$_5$. Based on the consistency in the location of the diffraction rings, we hypothesize that the crystalline features that form exhibit this phase. (c) Precipitate map produced by placing the virtual apertures seen in (a). We observe that the precipitates in this oxide are between 1-3 nm in lateral dimension.}
\label{fgr:S3}
\end{figure*}

\end{document}